\documentclass[useAMS,usenatbib]{mn2e}
\usepackage{graphicx}
\usepackage{subfigure}
\usepackage{amsmath}
\usepackage{cancel}

\title[The theory of globulettes]{The theory of globulettes: candidate precursors of brown dwarfs and free floating planets in H\,\textsc{II} regions}
\author[T. J. Haworth et al ]{Thomas J. Haworth$$\thanks{E-mail:
thaworth@ast.cam.ac.uk}, Stefano Facchini and Cathie J. Clarke  \\
Institute of Astronomy, Madingley Road, Cambridge, CB3 0HA}
\begin{document}

\date{Accepted ???. Received ???; in original form ???}

\pagerange{\pageref{firstpage}--\pageref{lastpage}} \pubyear{2014}

\maketitle
\label{firstpage}

\begin{abstract}
Large numbers of small opaque dust clouds - termed `globulettes' by Gahm et al - have been observed in the H\,\textsc{ii} regions surrounding young stellar clusters. With masses typically in the planetary (or low mass brown dwarf) regime, these objects are so numerous in some regions (e.g. the Rosette) that, if only a small fraction of them could ultimately collapse, then they would be a very significant source of free floating planets. Here we review the properties of globulettes and present a theoretical framework for their structure and evolution. We demonstrate that their interior structure is well described by a pressure confined isothermal Bonnor-Ebert sphere and that the observed mass--radius relation ($M\propto R^{2.2}$) is a systematic consequence of a column density threshold below which components of the globulette are not identified. We also find that globulettes with this interior structure are very stable against collapse within H\,\textsc{ii} regions. {We follow Gahm et al in assuming that globulettes are detached from the tips of pillars protruding in from the swept up shell that borders the expanding H\,\textsc{ii} region and produce a model for their dynamics, finding that globulettes will eventually impact the shell. We derive an expression for the time it takes to do so and show that dissipation of energy via dust cooling allows \textit{all} globulettes to survive this encounter and escape into the wider ISM. Once there the ambient pressure drops and they disperse on timescales around 30--300 kyr and should be observable using ALMA out to distances of order a parsec. Since we find that globulettes are stable, the only route via which they might still form brown dwarfs or planets is during their collision with the shell or some other violent perturbative event.}

\end{abstract}

\begin{keywords}
ISM: H\,{\textsc{ii}} regions -- ISM: clouds -- ISM: kinematics and dynamics --  ISM: Bubbles  -- planets and satellites: formation -- (stars:) brown dwarfs

\end{keywords}

\section{Introduction}
\label{introduction}
Globulettes are small, dense conglomerations of gas observed within H\,\textsc{ii} regions, recently brought to attention by \cite{2007AJ....133.1795G}, \cite{2013A&A...555A..57G}, \cite{2014A&A...565A.107G} and \cite{2014A&A...567A.108M}. \cite{2014A&A...565A.107G} studied globulettes in the Carina nebula, finding that they are typically approximately spherical, having average radii varying from about 0.1 to 10\,kAU, masses from 0.1 to 100 Jupiter masses and number densities in the range $10^{3}-10^{5}$\,cm$^{-3}$. They also noted that the globulette masses and radii were found to be statistically correlated.

There are two possible formation mechanisms for globulettes. H\,\textsc{ii} regions are the expanding bubbles of ionised gas about massive stars. As H\,\textsc{ii} regions expand, they accumulate material in a shell about their periphery. In the first globulette formation mechanism they are the detached tips, or fragmented remnants, of pillars (a.k.a elephant trunks) which are known to reside at the periphery of H\,\textsc{ii} regions and to be moving at the same velocity as the expanding shell. Once detached the globulettes subsequently move in the same direction as the expanding shell. This is the formation mechanism proposed by \cite{2007AJ....133.1795G} and \cite{2013A&A...555A..57G}. The justification is that globulettes are often in the vicinity of pillars and also that globulette velocities are observed to be similar to that of the shell \citep{2013A&A...555A..57G}. 

\cite{2012A&A...546A..33T} also showed that larger globules (a few solar masses - though these sizes were limited by the spatial resolution of the simulation) within the H\,\textsc{ii} region appear following the irradiation of a turbulent medium.  As well as being more massive, these objects are also more complicated in structure, for example being elongated with tails, similar to some of the globulettes studied by  \cite{2007AJ....133.1795G} and \cite{2014A&A...565A.107G}.  In this latter case the globule has a more random motion determined by the initial turbulent velocity field. 

\cite{2007AJ....133.1795G} showed that the expected photoevaporation timescale
for globulettes is around $3.8$ Myr; this value is consistent with
the fact that images of globulettes show a mixture of spherical and
cometary morphologies. \cite{2014A&A...565A.107G} found that some globulettes in Carina were particularly small and dense, leading the authors to conclude that perhaps they were at a later stage in globulette evolution, having had more time to be compressed or for lower density outer layers to be dispersed. They also speculate that globulettes might form free-floating planets or brown dwarfs, though considerable external pressure may be required to collapse globulettes. This external pressure could be in the form of turbulence, photoevaporative driving or ram pressure \citep{2007AJ....133.1795G}. 

In this paper we  develop a model to describe the internal structure, motions and stability of globulettes in (and exterior to) H\,\textsc{ii} regions. We aim to determine if/when a globulette will have to pass through the shell bounding the H\,\textsc{ii} region and whether it will survive the encounter. We also aim to understand the sensitivity of globulette lifetimes to star forming region properties. In doing so we shall help to constrain models of globulette formation and any possible subsequent evolution into brown dwarfs or vagrant planets.

\section{Internal structure and the mass--radius relation}
\label{struct}
We fitted the masses and radii of globulettes in the Carina nebula observed by \cite{2014A&A...565A.107G} and found that  the mass--radius relation differs from the cubic power law expected for constant density spherical clouds, rather being given by 
\begin{equation}
	\left(\frac{M}{M_\textrm{J}}\right) = 1.6\times\left(\frac{R_{\textrm{G}}}{\textrm{kAU}}\right)^{2.2}
	\label{MR}
\end{equation}
to understand this we require a model of the internal structure of globulettes.

\subsection{Globulettes as Bonnor-Ebert spheres}
\label{globBES}
Using CO observations of small globules (before the term globulette was coined) \citet{1994ApJ...430L.125G} concluded that globulettes are isothermal since they are optically thick to FUV radiation from the nearby stars. More recently, \citet{2013A&A...555A..57G} have shown that these objects might have a hot thin envelope surrounding a cold dense core, as suggested by $^{12}$CO emission lines. As a first approximation, an isothermal sphere is still pragmatic. If we further assume hydrostatic equilibrium, we can model them as Bonnor-Ebert spheres, the density of which is described by the Lane--Emden equation


\begin{equation}
\label{eq:be_spheres}
\frac{1}{R^2}\frac{d}{dR}\big( R^2\frac{d}{dR}\ln (\rho/\rho_{\textrm{c}})\big) = -\frac{4\pi G\rho_{\textrm{c}}}{c_{\rm s}^2}\exp[\ln (\rho/\rho_{\textrm{c}})],
\end{equation}
where $R$ is the radial coordinate, $c_{\rm s}$ is the sound speed, $\rho$ is mass density and $\rho_{\textrm{c}}$ is the density value at the {centre of the sphere}. We define $R_0=c_{\rm s}/(4\pi G \rho_{\textrm{c}})^{1/2}$ as the typical length scale of the problem. By substituting $\psi = \ln(\rho/\rho_{\textrm{c}})$ and $x=R/R_0$, we can rewrite equation \ref{eq:be_spheres} as:

\begin{equation}
\frac{1}{x^2}\frac{d}{dx}\big(x^2\frac{d}{dx}\psi\big)=-\exp{\psi}.
\end{equation}
We numerically solve this second order differential equation by solving the two coupled first order equations:

\begin{equation}
\frac{d\psi}{dx}=\frac{y}{x^2},
\end{equation}

\begin{equation}
\frac{dy}{dx}=-x^2\exp\psi.
\end{equation}


We use a standard ODE integrator where the grid over $x$ is logarithmically distributed between $10^{-4}$ and $10^4$ with $800$ points to solve for the radial density structure. In order to solve the equation we require that $\psi(x=0)=0$ and $y(x=0)=0$. 

\begin{figure}
\center
\includegraphics[width=\columnwidth]{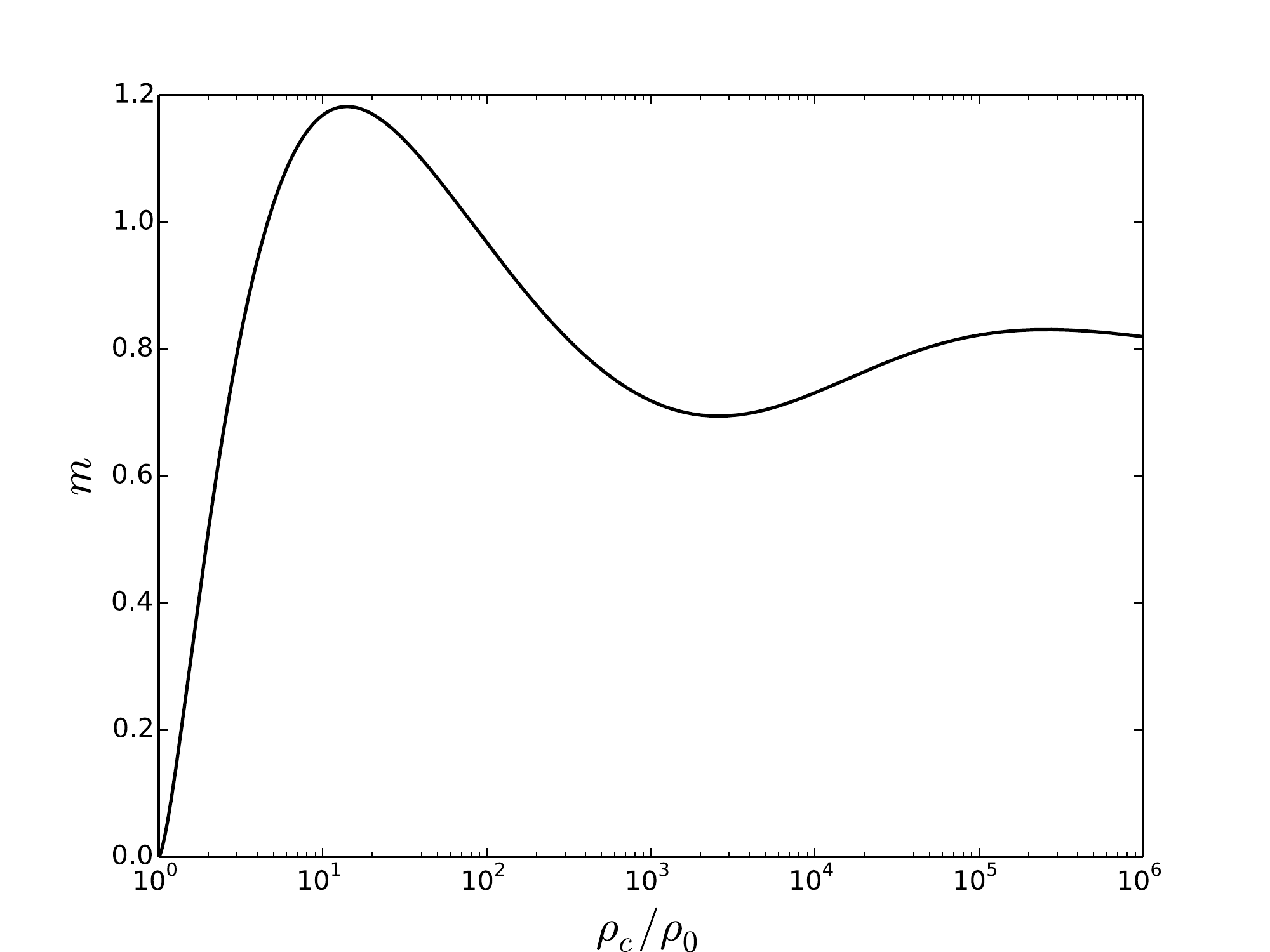}
\caption{Dimensionless mass $m$ versus $\rho_c/\rho_0$. This relation is used to obtain $\rho_c$ for a given $M$, $P_{\rm ext}$ and internal temperature. If $m$ is less than $0.6$, the solution for $\rho_c/\rho_0$ is unique, and the sphere is stable against contraction.}
\label{fig:m_rho}
\end{figure}

In order to define the unique solution of the dimensional density $\rho(R)$ for a given mass, we need to determine two physical parameters: $\rho_c$ and $R_{\rm tr}$, i.e. the radius at which we truncate the sphere \citep[see][for a further description of the technique we are using]{2005fost.book.....S}. By obtaining these two parameters, we can uniquely relate a mass to a radius. The two known parameters for the globulettes are: the internal temperature $T$, and the external pressure $P_{\rm ext}$ of the H\,{\textsc{ii}} region. The external pressure is related to the density of the sphere $\rho_0$ at the outer radius $R_{\rm tr}$ via the simple relation $\rho_0=P_{\rm ext}/c_{\rm s}^2$, since the internal pressure at the outer edge of the cold neutral gas has to match the external pressure of the less dense, much hotter ionized gas of the H\,{\textsc{ii}} region. We use an internal temperature of { 10 K,  a canonical value for dense opaque clouds. \citet{1994ApJ...430L.125G} estimated a temperature of the globulettes in the Rosette of $\sim 15$\,K from the CO emission line width. We assume} a mean molecular weight $\mu=2.3$. For the H\,\textsc{ii} region properties we assume a temperature of $10^4$\,K and a number density of $10$\,cm$^{-3}$. We calculate globulette properties over a logarithmically spaced mass distribution between $10^{-1}$ and $10^3$ $M_J$.

\begin{figure*}
\center
\includegraphics[width=14cm]{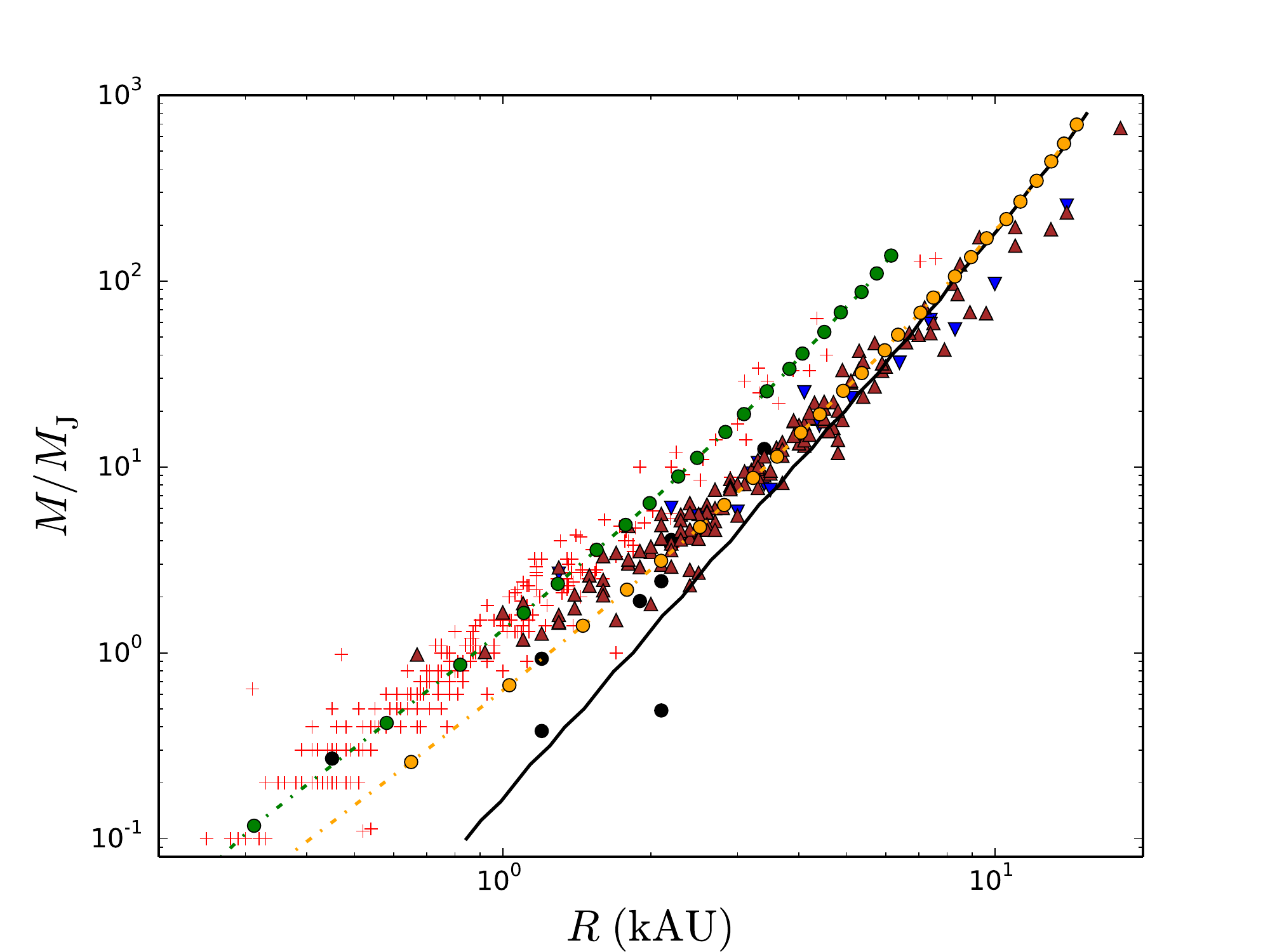}
\caption{The mass--radius relationship from our theoretical models and from observations. The back solid line indicated the mass-radius relation estimated by assuming the globulettes are isothermal Bonnor-Ebert spheres. The yellow dashed-dotted line (with circled markers) indicate the mass-radius relation { in Rosette} obtained after accounting for systematic errors in the mass determination, by truncating the spheres to the point where their column density equals the detectability threshold. { The green dashed-dotted line (with circled markers) shows the same result for Carina.} Brown triangles are globulettes in the Rosette, red crosses are globulettes in the Carina nebula \citep{2014A&A...565A.107G}, blue triangles in IC 1805 and black circles in NGC 7822 \citep{2007AJ....133.1795G}.  }
\label{fig:mass_radius_relation}
\end{figure*}

In order to find $\rho_c$, we firstly define a dimensionless mass:

\begin{equation}
m=\frac{P_{\rm ext}^{1/2}G^{3/2}M}{c_{\rm s}^4}.
\end{equation}
The dimensionless mass $m$ can be numerically related to $\rho_c/\rho_0$, as shown in Figure \ref{fig:m_rho}, via the following relation:

\begin{equation}
m = \big( 4\pi\frac{\rho_c}{\rho_0} \big)^{-1/2} \big( x^2 \frac{d\psi}{dx} \big).
\end{equation}
Therefore from a given mass, external pressure, and internal temperature of a globulette, we are able to extract the value of $\rho_c$. Finally, we integrate outwards the density function until the total mass matches with the given one. We can therefore determine the value of the truncation radius $R_{\rm tr}$. Doing so over a range of masses yields the theoretical mass-radius relation for globulettes. 

In Figure \ref{fig:mass_radius_relation} the solid black line shows {our theoretical} mass-radius relation for an external pressure $P_{\rm ext}=n k_B T$, where the { electron} density $n$ in the H\,{\textsc{ii}} region is $10$\,cm$^{-3}$, and the { average} temperature of the ionized gas is $5800$\,K { \citep[that estimated for the Rosette by][]{1985A&A...144..171C}}. Overlaid are mass and radii of globulettes from the Rosette  \citep[brown triangles, from][]{2007AJ....133.1795G},  Carina \citep[red crosses, from][]{2014A&A...565A.107G}, IC 1805 \citep[blue triangles, from][]{2007AJ....133.1795G} and NGC 7822 \citep[black circles, from][]{2007AJ....133.1795G}. These pressure confined Bonnor-Ebert spheres are not very stratified, having a broad interior region of slowly varying density until close to the globulette boundary (c.f. Figure \ref{fig:ratio_rhos}). The theoretical mass--radius relation is well represented by a power-law, where $M\propto R^3$ (as one would expect) and the high mass regime of the data seems to be quite well described by this distribution (i.e. they are Bonnor-Ebert spheres). However, in the low mass regime the discrepancy between data and model is prominent (more than an order of magnitude). We now demonstrate that this can be explained by observational systematics.

\subsection{Correcting for observational systematics}

So far we have found that our model for the internal structure of globulettes describes well the observed mass--radius relation of large radius globulettes { in the Rosette nebula}, but masses differ by over an order of magnitude for smaller globulettes. \citet{2007AJ....133.1795G} estimated globulette masses using extinction maps derived from deep narrowband H$\alpha$ images, collected with the 2.6 m Nordic Optical Telescope \citep{2007AJ....133.1795G}. { For Carina, \citet{2014A&A...565A.107G} used H$\alpha$ data from the Hubble Space Telescope archive.} In order to estimate the masses, { in both cases} they measure the column density of the neutral cold gas of the globulettes. By using this method, there has to be an intrinsic threshold, under which no column density is detected via extinction. From \citet{2007AJ....133.1795G}, we estimate this threshold to be $N_{\rm th} \sim 4\times 10^{20}$\,cm$^{-2}$ { in the Rosette}.

We construct an \emph{observed} mass-radius relation  by using the following procedure. From the theoretical mass-radius relation, we compute the surface density profile of the spheres in the plane of the sky. The relation between the two can be expressed by \citep[e.g.][]{2009MNRAS.395.1092D}:

\begin{equation}
\Sigma(r)=2\int_r^R{\frac{\rho(R)RdR}{\sqrt{R^2-r^2}}},
\end{equation}
where $\Sigma(r) = N(r)/\mu m_H$ and $r$ is the cylindrical radius of the sphere in the plane of the sky. We have checked that we re-obtain the initial mass, by integrating the surface density profile out to a cylindrical radius equalling $R_{\rm tr}$:
\begin{equation}
M = \int_0^{R_{\rm tr}}{\Sigma(r)2\pi rdr}.
\end{equation}

We then truncate the projected spheres to a cylindrical radius $r_{\rm th}$ such that $N(r_{\rm th}) = N_{\rm th}$. This is the outer radius of the \emph{observed} globulettes. We finally compute the new mass of the globulette $M_\Sigma$ (since it comes from the surface density) by integrating the surface density profile from the center out to $r_{\rm th}$.  The new mass-radius relation is given by $M_\Sigma(r_{\rm th})$, and is shown with the yellow dashed-dotted line in Figure \ref{fig:mass_radius_relation}. A much better agreement between the data of the Rosette nebula and the model is apparent. The yellow circles on the line represent the points through which the line is drawn. The points are not equally separated in logarithmic space (though our logarithmic sampling of masses was, c.f. section \ref{globBES}). This is indicating that the low mass end of the distribution comes from initial masses that are higher by almost an order of magnitude, i.e. the low mass regime is probing the cores of globulettes with masses equal to a few $M_J$.  { This same mass-radius distribution also describes the data points from IC 1805 well, indicating that the two star forming regions have similar external pressures. The data points for NGC 7822 are so few and the scatter is so large that we cannot make any similar comment on the relative conditions there.

We also produce a theoretical mass--radius relation for Carina. The average electron density is well observationally constrained at $30$\,cm$^{-3}$ from Ne\,\textsc{ii} emission line data \citep{2006ApJ...652L.125O,2011ApJ...739..100O}. We keep the same values used before for the internal temperature of globulettes and temperature of the ionised gas, respectively $10$ and $5800$\,K. However, in order to obtain a good agreement with the data, we require a higher column density threshold, namely $N_{\rm th}=8\times10^{20}$. We cannot estimate this new column density threshold directly from \citet{2014A&A...565A.107G}; however we expect $N_{\rm th}$ to be higher in Carina than in Rosette, since Carina is further away, at $\sim 2.3$\,kpc \citep[e.g.][]{2006ApJ...644.1151S}, whereas Rosette is at $\sim1.6$\,kpc \citep[e.g.][]{2002AJ....123..892P}. Moreover, in Carina there is potentially more foreground emission due to the higher electron density, therefore decreasing the signal to noise of the extinction maps. By assuming the aforementioned value for $N_{\rm th}$, the agreement between the model and the observational data is good, as shown by the green dashed-dotted line and circles in Figure \ref{fig:mass_radius_relation}.}

We conclude that globulettes seem to be well described by isothermal Bonnor-Ebert spheres of low density stratification. However, in order to properly compare this underlying model with the data, we need to account for systematics due to the observational techniques used to estimate the globulettes masses and radii.




\section{Stability against collapse}
\label{stability}
\begin{figure}
\center
\includegraphics[width=\columnwidth]{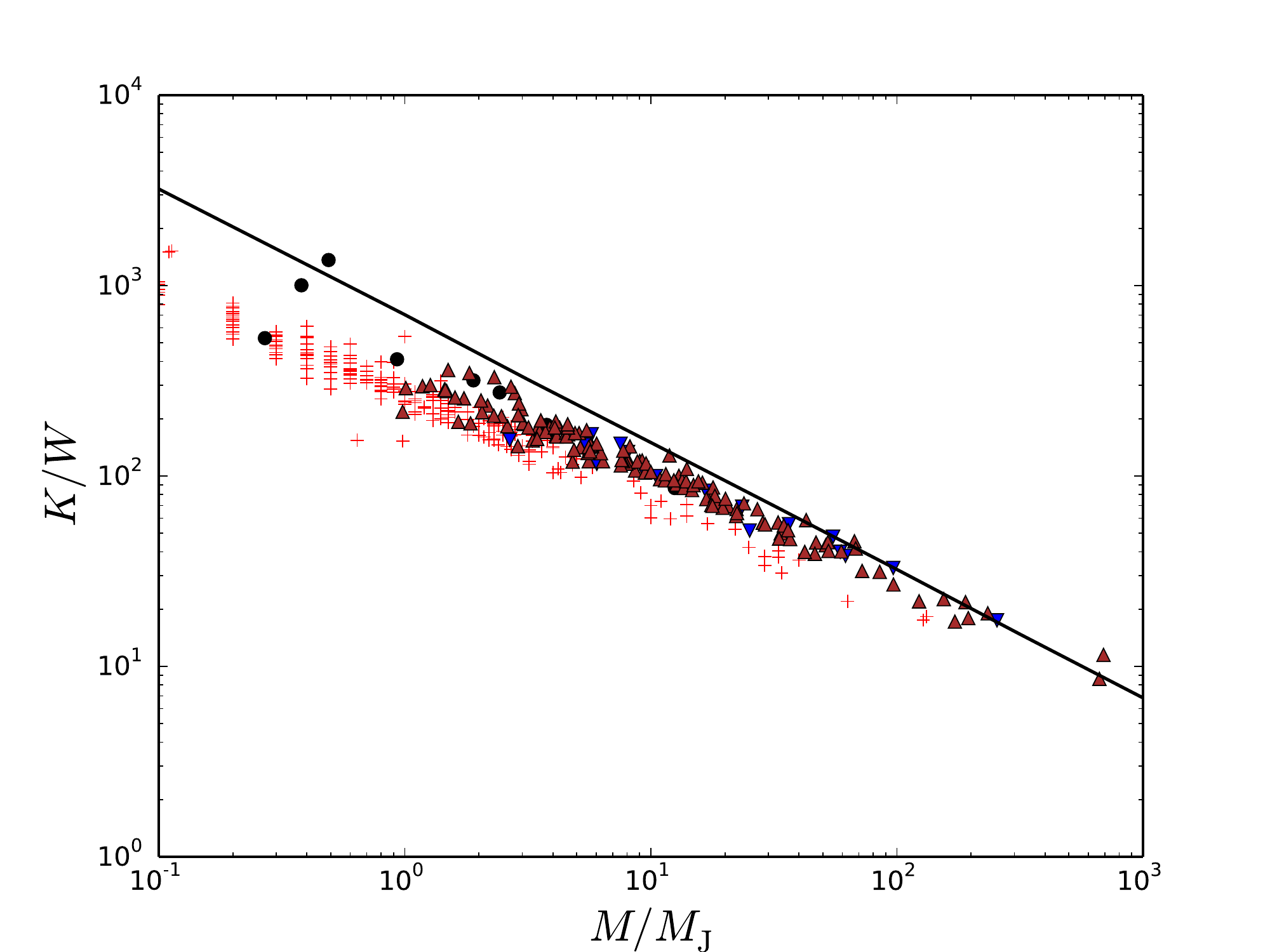}
\caption{The ratio of thermal and binding energy of the globulettes, as a function of mass. The points represent globulettes in the same systems as those with corresponding symbols in Figure \ref{fig:mass_radius_relation} and the line is the ratio for our model of globulette structure. This ratio is always much higher than $1$, indicating that these structures are not gravitationally bound. They are confined by external pressure from the hot ionized medium of the H\,{\textsc{ii}} region.}
\label{fig:stability}
\end{figure}

\begin{figure}
\center
\includegraphics[width=\columnwidth]{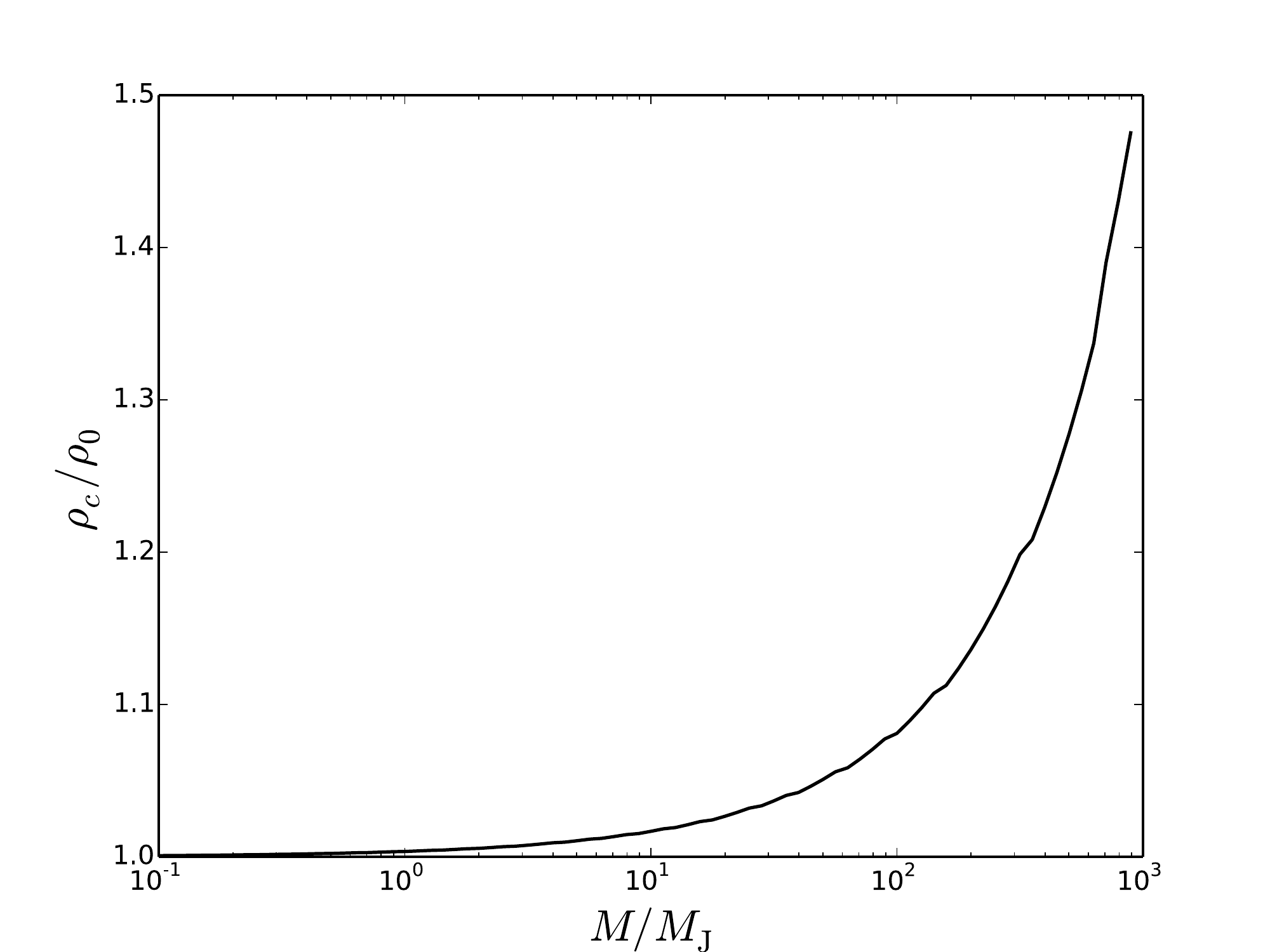}
\caption{Ratio of the central density over density at the outer edge of globulettes versus mass. Since $\rho_c/\rho_0$ is well below the value $14.1$, these spheres are stable against contraction.}
\label{fig:ratio_rhos}
\end{figure}

We now discuss whether globulettes are stable against contraction or not. In Figure \ref{fig:stability} we show the ratio of internal (thermal) energy $K$ to the absolute value of gravitational energy $W$ for the globulettes considered in this paper. Since at first order globulettes are isothermal and characterized by a uniform density, this ratio can be evaluated via the simple relation:

\begin{equation}
\frac{K}{W}=\frac{5}{2}\frac{c_{\rm s}^2R}{GM}.
\end{equation}
The thermal energy is at least by an order of magnitude greater than the binding energy. Globulettes are not gravitationally bound structures. They are pressure-confined by the hot surrounding gas. A natural question is whether the pressure in the H\,{\textsc{ii}} region is high enough to trigger contraction in the globulettes. It is well known that Bonnor-Ebert spheres are stable against contraction when $\rho_c/\rho_0<14.1$ \citep[see e.g.][]{2005fost.book.....S}.  For the mass-radius relation estimated in Section \ref{struct}, $\rho_c/\rho_0$ is always less than $1.5$, as shown in Figure \ref{fig:ratio_rhos}. Therefore, such globulettes are stable against contraction in a typical H\,{\textsc{ii}} region. In order to contract, they would need either a higher external pressure or a higher mass. We can conclude that globulettes will not collapse in the H\,{\textsc{ii}} region, unless they are subject to very high pressures (much higher than usual thermal and turbulent pressures). They are therefore very unlikely to form brown dwarfs or free-floating planets, unless contraction is triggered by some strong perturbation, such as colliding with the shell of the H\,{\textsc{ii}} region.

\section{H\,\textsc{ii} region confinement timescale}
\subsection{Analytic model}
\label{anal}
We build our model based upon the conclusions drawn from observations discussed in the introduction \citep[e.g. by][]{2007AJ....133.1795G}. The picture we have is that there is an expanding H\,\textsc{ii} region, bounded by a dense shell, from which pillars protrude into the ionised gas. 

There are already well established simple models describing the extent of an H\,\textsc{ii} region with time.  An ionising source in a uniform density hydrogen medium rapidly ionises a sphere of Str\"omgren radius given by
\begin{equation}
	r_{\textrm{s}} = \left(\frac{3N_{\textrm{ly}}}{4\pi n_{\textrm{e}}^2 \alpha_{\textrm{B}}}\right)^{1/3}
\end{equation}
where $N_{\textrm{ly}}$, $n_{\textrm{e}}$ and $\alpha_{\textrm{B}}$ are the number of ionising photons emitted per second, electron density in the H\,\textsc{ii} region and case B recombination coefficient for hydrogen. Up to the Str\"omgren radius the ionisation front propagation is rapid and does not significantly alter the density distribution. 

The subsequent expansion of the H\,\textsc{ii} region is known as D-type and results in the sweeping up of  a shell at the boundary of the H\,\textsc{ii} region. \cite{2006ApJ...646..240H} solved the equation of motion of the shell to find its location as a function of time, which is essentially coincident with the ionisation front given that the shell is thin 
\begin{equation}
	r_{\textrm{I}}(t) = r_{\textrm{s}}\left(1 + \frac{7\sqrt{4}\, c_{\textrm{I}} t}{4\sqrt{3}\, r_{\textrm{s}}}\right)^{4/7}
	\label{hi}
\end{equation}
where $c_{\textrm{I}}$ is the sound speed in the ionised gas. Equation \ref{hi} gives slightly faster D--type expansion than
the classic solution by \cite{1998ppim.book.....S} since it is based on momentum conservation
rather than a direct linkage of the pressure in the HII region to the ram
pressure in the ambient medium: the fact that these assumptions give
slightly different relations is an indication that the thin-shell {approximation, which both approaches use,}
is not accurate in practice: see \cite{2012MNRAS.419L..39R}.
We only consider the D-type expansion of H\,\textsc{ii} regions since it is in this phase that pillars which are dynamically associated with the shell could  form.


At ``detach'' time $t_{\textrm{D}}$ a globulette detaches from a pillar of length $L_{\textrm{p}}$ when the shell is at position $r_{\textrm{D}}$. We assume that the properties (mass, radius) of the globulette do not change with time. We also ignore any acceleration due to the rocket effect or direct radiation pressure, the impact of which we discuss in section \ref{uncertainty}. In the absence of ram pressure  acting upon the globulette (we will later demonstrate that ram pressure has no effect on the globulette propagation) the subsequent location of the {globulette} as a function of time is 
\begin{equation}
	r_{\textrm{g}}(t) = r_{\textrm{D}} - L_{\textrm{p}} + \dot{r_{\textrm{I}}}(t_{\textrm{D}}) t.
	\label{globpos}
\end{equation}
In the event that $r_{\textrm{g}}(t)$ and $r_{\textrm{I}}(t)$ are equal at some time $t_{\textrm{c}}$, which we call the confinement timescale, the globulette will collide with the shell. However equating equations \ref{hi} and \ref{globpos} yields an implicit relation for which the globulette lifetime needs to be solved numerically. 

We can derive an explicit expression for the globulette confinement timescale in the limit of constant shell deceleration (i.e. for small pillar sizes).  In this limit of constant shell deceleration the shell position following the detachment of a globulette is given by
\begin{equation}
	r_{\textrm{I}}(t) = r_{\textrm{D}} +  \dot{r_{\textrm{I}}}(t_{\textrm{D}}) t + \frac{1}{2}\ddot{r_{\textrm{I}}}(t_{\textrm{D}}) t^2.
	\label{shellpos}
\end{equation} 
At time $t_{\textrm{C}}$ the globulette collides with the shell when equations \ref{globpos} and \ref{shellpos} are equal, i.e.
\begin{equation}
	t_{\textrm{C}}^2 = -\frac{2L_{\textrm{p}}}{\ddot{r_{\textrm{I}}}(t_{\textrm{D}})}.
	\label{lp1}
\end{equation}
Rewriting equation \ref{hi} as
\begin{equation}
	r_{\textrm{I}} = r_{\textrm{s}}\left(1 + \frac{t}{t_{\textrm{o}}}\right)^{4/7}
\end{equation}
then
\begin{equation}
	\ddot{r_{\textrm{I}}}(t_{\textrm{D}}) = -\frac{12}{49}\frac{r_{\textrm{s}}}{t_{\textrm{o}}^2}\left(1+\frac{t_{\textrm{D}}}{t_{\textrm{o}}}\right)^{-10//7}.
\end{equation}
Substituting this into equation \ref{lp1}
\begin{equation}
	t_{\textrm{C}} = \sqrt{\frac{49L_{\textrm{p}}t_{\textrm{o}}^2(1+t_{\textrm{D}}/t_{\textrm{o}})^{10/7}}{6r_{\textrm{s}}}}.
	\label{tl1}
\end{equation}
which can be written as
\begin{equation}
	t_{\textrm{C}} = \sqrt{\frac{49L_{\textrm{p}}}{6r_{\textrm{D}}}}\left(1+\frac{t_{\textrm{D}}}{t_{\textrm{o}}}\right)t_{\textrm{o}}.
	\label{tl2}
\end{equation}
where $r_D$ is the ionisation front position at the time of globulette detachment.
If the size of pillars scales linearly with the size of the H\,\textsc{ii} region $L_{\textrm{p}}=fr_{\textrm{D}}$ then the lifetime scales linearly with detach time. However if the pillar size is a constant then 
\begin{equation}
	t_{\textrm{C}} = \sqrt{\frac{49L_{\textrm{p}}}{6r_{\textrm{s}}}}\left(1+\frac{t_{\textrm{D}}}{t_{\textrm{o}}}\right)^{5/7}t_{\textrm{o}}
	\label{lifeequn}
\end{equation}
so confinement timescale scales as $t_D^{5/7}$. 

The key points to note are that {under the assumptions that we have made} the globulette lifetime only depends upon the time at which they are detached and the length of the pillar from which they are detached. It is insensitive to the source ionising flux since  $t_{\textrm{C}}\propto N_{\textrm{ly}}^{-1/6}$. It is also independent of the globulette properties (though this would not be the case if ram pressure were important). {Given that we have not included the rocket effect the sensitivity to ionising flux will be stronger than the weak dependence quoted above, we discuss this in more detail in section \ref{uncertainty}.}

\begin{figure}
	\includegraphics[width=8.cm]{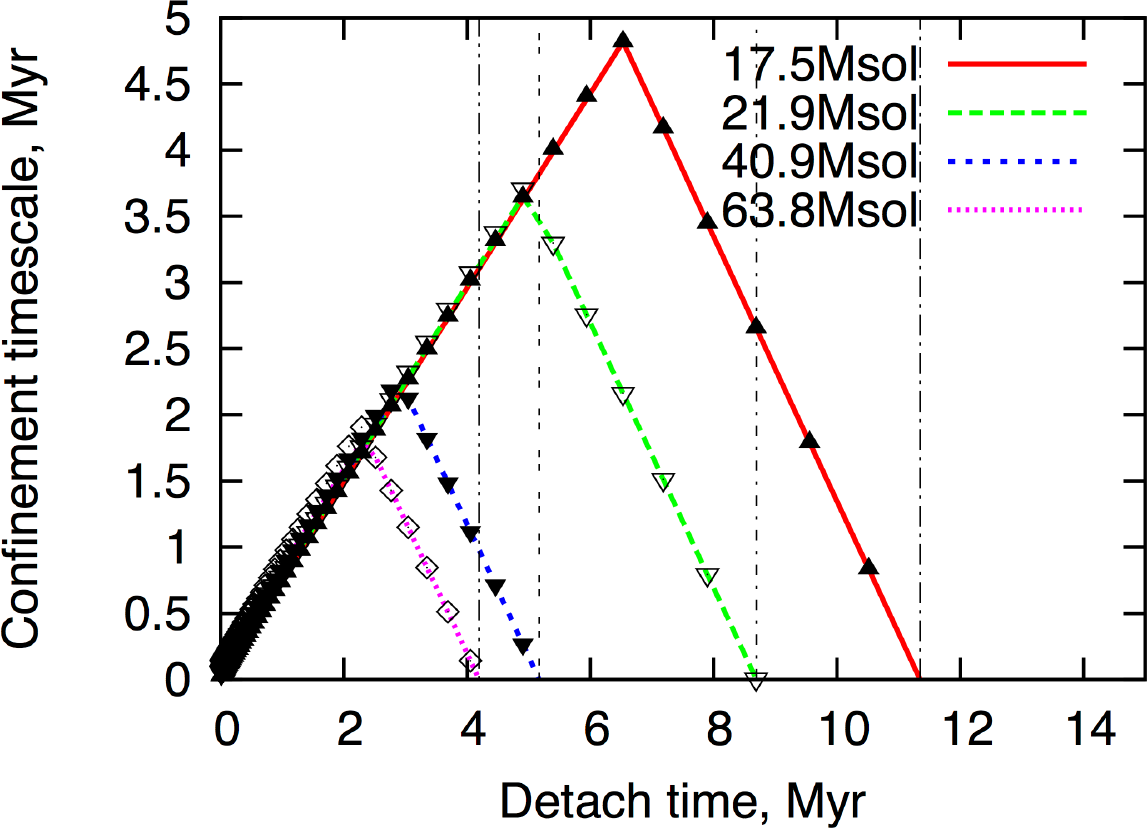}
	\includegraphics[width=8.cm]{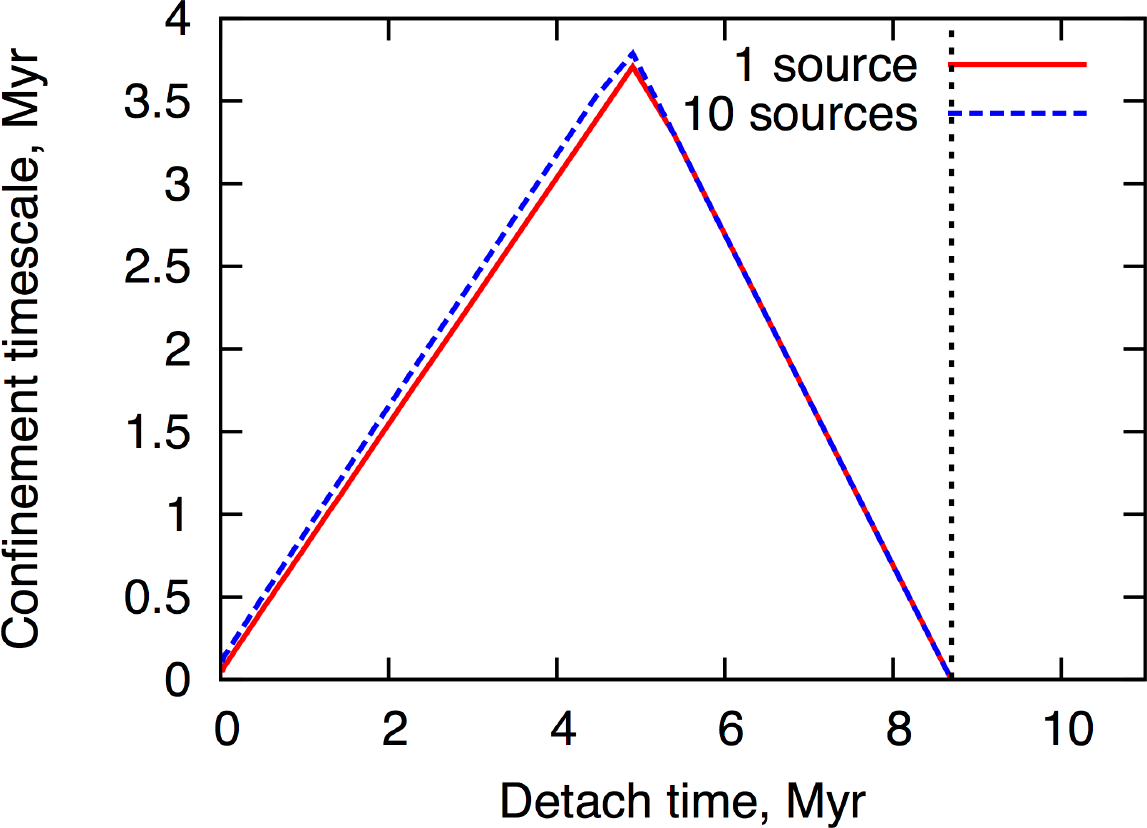}
	\hspace{10pt}
	\includegraphics[width=8.cm]{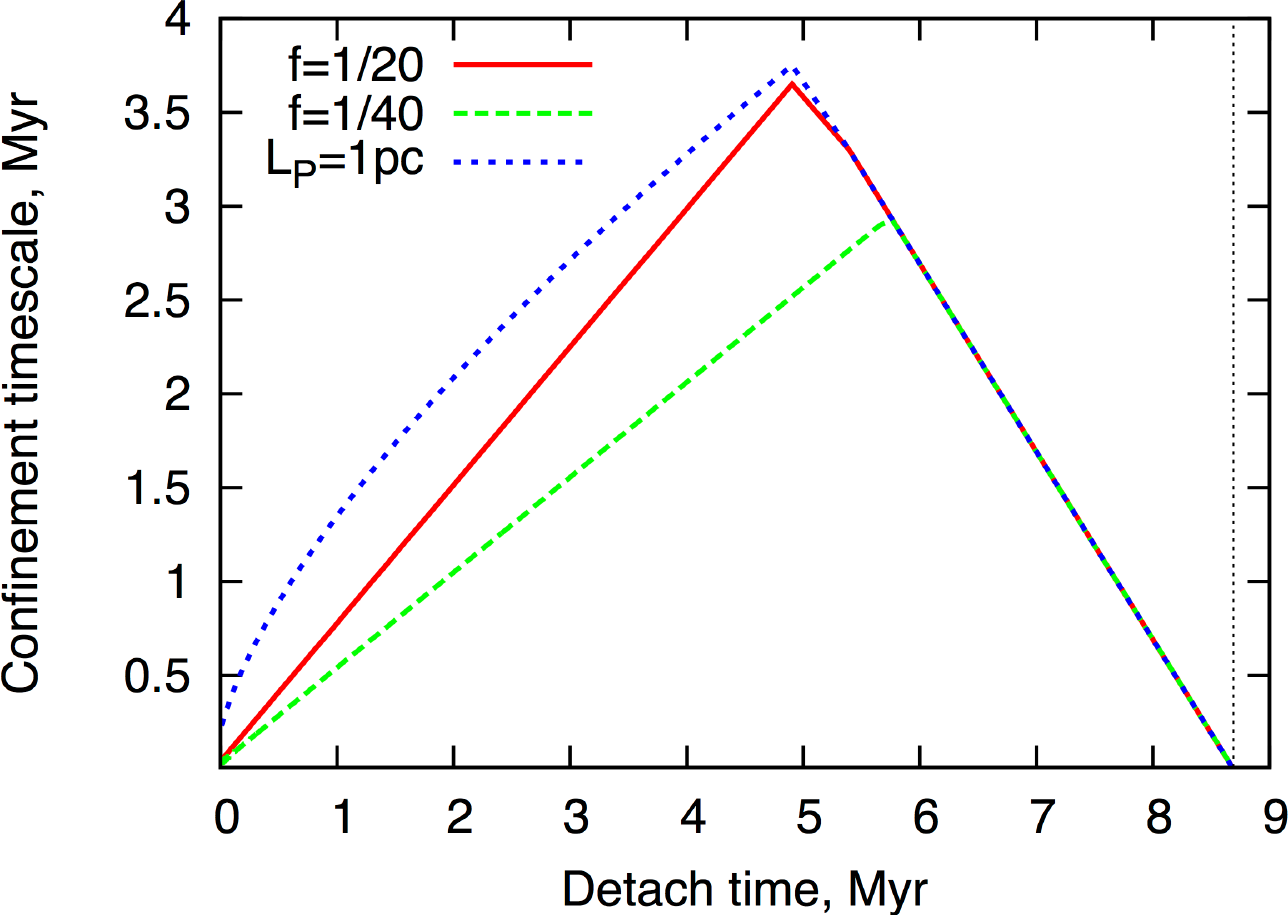}
		\caption{The time following detachment from the shell that a globulette remains confined to an H\,\textsc{ii} region. The top panel shows the result for different stellar masses and both with (lines) and without (points) ram pressure. The middle panel demonstrates that the result is insensitive to the ionising flux and the bottom right panel shows results for models with different pillar size prescriptions. Vertical lines denote the star lifetime. }
	\label{lifetimes}
\end{figure}



\subsection{Numerical comparison}
\label{numstuff}

\label{numcom}
We now compare our analytic model with more detailed numerical calculations. This allows us to test the effect of ram pressure, which was neglected in our analytic model. 

\subsubsection{Shell and globulette evolution}
The expansion of the shell is modelled using Equation \ref{hi}. As in our analytic model, we assume that globulettes are detached from pillars suspended from the shell at some time after the start of D-type expansion. The globulettes are spherical and their initial velocity is the same as that of the shell.

Following detachment, the globulette is assumed to be impeded by ram pressure only and so the velocity evolves according to
\begin{equation}
	\frac{\textrm{d}v}{\textrm{d}t} = -\frac{\rho_{\textrm{I}}  \pi R_{\textrm{G}}^2 \left(v - v_{\textrm{H}_{\textrm{II}}}\right)^2}{M_{\textrm{G}}}
	\label{du}
\end{equation}
where $R_{\textrm{G}}, M_{\textrm{G}}, \rho_{\textrm{I}}$ and $v_{\textrm{H}_{\textrm{II}}}$ are the globulette radius and mass and the density and velocity of the ionised gas in the radial direction (i.e. the globulette propagation direction). 

The velocity in the ionised gas in the immediate vicinity of the globulette evolves according to
\begin{equation}
	v_{\textrm{H}_{\textrm{II}}} = \frac{1}{2}\left(\frac{r_{\textrm{G}}}{r_{\textrm{I}}}\right)\dot{r_{\textrm{I}}}
	\label{ram}
\end{equation}
where $r_{\textrm{G}}$, $r_{\textrm{I}}$ and $\dot{r_{\textrm{I}}}$ are the globulette position, ionisation front position and ionisation front velocity. The density in the ionised gas also evolves with time since ionisation equilibrium imposes that
\begin{equation}
	\rho_{\textrm{I}}^2 r_{\textrm{I}}^3 = \textrm{constant}
\end{equation}
and so 
\begin{equation}
	\rho_{\textrm{I}} = \rho_{\textrm{a}}\sqrt{\frac{ r_{\textrm{s}}^3}{r_{\textrm{I}}^3}}.
	\label{rhoevolution}
\end{equation}
where $\rho_{\textrm{a}}$ is the ambient density.

\subsubsection{Further implementation}
We begin by updating the ionisation front position in regular time intervals $\textrm{d}t=t_*/\left(4\times10^5\right)$ where $t_*$ is the lifetime of the ionising source. We use ionising fluxes from \cite{1998ApJ...501..192D} for solar metallicity stars of 17.5, 21.9, 40.9 and 63.8\,M$_{\odot}$. These ionising fluxes were computed by integrating over LTE metal--line blanketed model atmospheres from \cite{1979ApJS...40....1K}. Lifetimes of the ionising sources are calculated using the giant branch and core helium burning timescale models from \cite{2000MNRAS.315..543H}. The stellar properties are summarised in Table \ref{starparams}. 

\begin{table}
 \centering
 \begin{minipage}{180mm}
  \caption{Properties of the ionising sources considered in this paper.}
  \label{starparams}
  \begin{tabular}{@{}l c l@{}}
  \hline
   Mass (M$_\odot$) & Ionising flux (photons/s)& Lifetime (Myr)\\
 \hline
   17.5 & $1.05\times10^{47}$ & 11.4 \\
   21.9 & $1.25\times10^{48}$ &  8.7 \\
   40.9 & $6.03\times10^{48}$ &  5.2\\
   63.8 & $2.24\times10^{49}$ &  4.2\\
\hline
\end{tabular}
\end{minipage}
\end{table}

At a time $t=t_{\textrm{D}}$ a globulette detaches from a pillar with velocity equal to that of the shell.
We then continue to evolve the shell and globulette positions numerically over time steps of ${\textrm{d}}t$ until $t_*$, beyond which the H\,\textsc{ii} region might soon be expected to be dispersed by supernovae. In the event of a collision with the shell (i.e. the globulette position overtakes the shell) we record the globulette confinement timescale $t_{\textrm{C}}=t-t_{\textrm{D}}$. In the event that $t_{\textrm{D}}+t_{\textrm{C}} > t_*$ then we set $t_{\textrm{C}} = t_*-t_{\textrm{D}}$ to account for the fact that the limited ionising source lifetime may cut short the globulette lifetime (or at least the time it is confined to the H\,\textsc{ii} region). Note that for the current study we assume that the globulette mass and radius remain unchanged once it is detached from the shell. 

\subsubsection{Results of numerical calculations}

Figure \ref{lifetimes} shows the globulette confinement timescale (the smaller of either the time until collision with the shell or the ionising source lifetime) as a function of detach time. The top panel shows the results for calculations with ionising sources of different masses. It also compares calculations with and without ram pressure included (lines and points respectively). Since there is no difference between the lines and the points we conclude that ram pressure has no effect on globulette evolution. We ran test calculations using ambient densities up to a factor $10^4$ higher (the ionised gas still obeyed equation \ref{rhoevolution}), but even in this extreme case the effect of ram pressure only elongated the globulette confinement time by at most 6 per cent. 

In the middle panel we also show that because $t_{\textrm{C}}\propto N_{\textrm{ly}}^{-1/6}$ (equation \ref{lifeequn}) that even if we increase the ionising flux by a factor of 10, the globulette confinement time is essentially unaffected.

\begin{figure}
	\includegraphics[width=8cm]{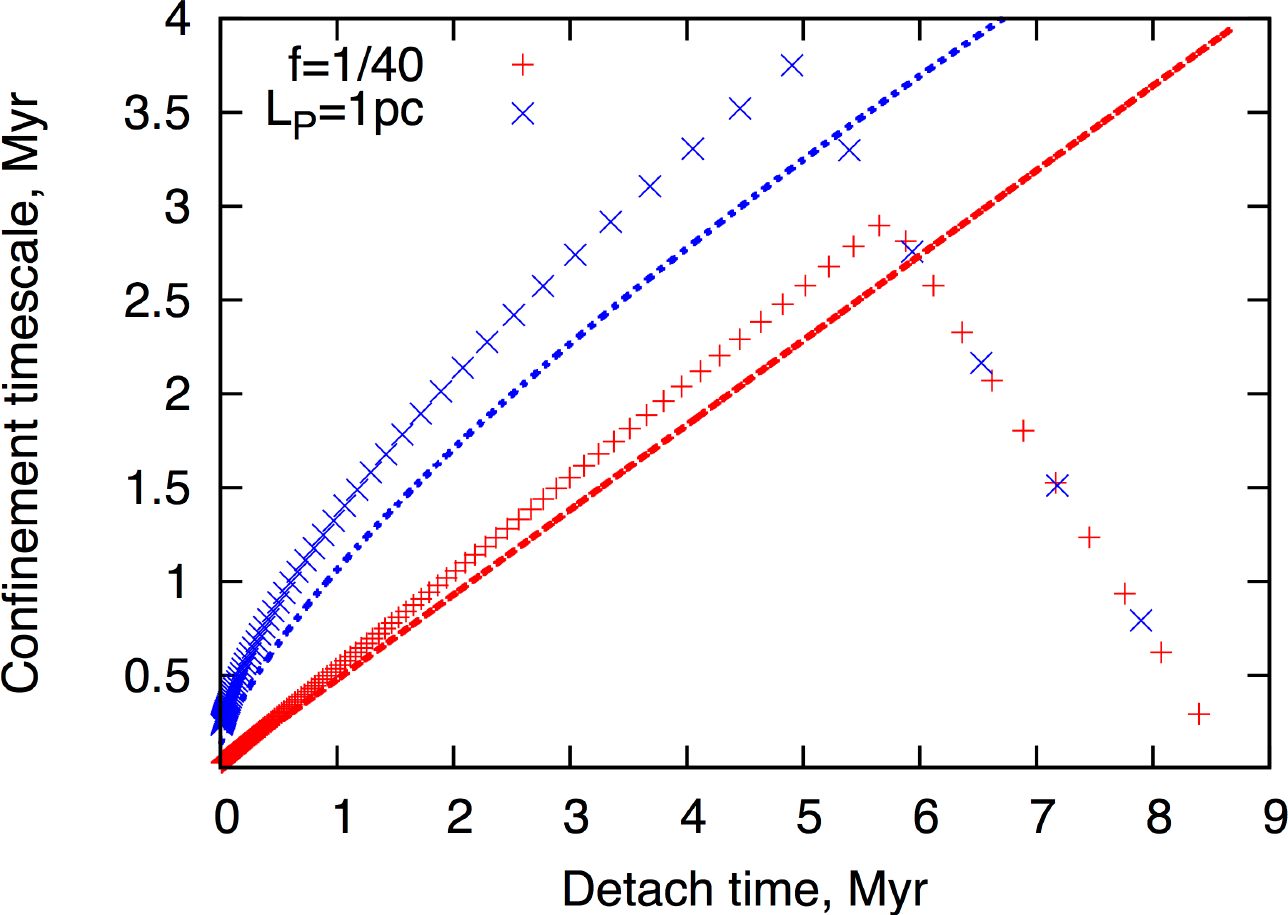}
		\caption{A comparison of our numerical calculations (points) with equation \ref{tl2} (lines) for different pillar size prescriptions. }
	\label{lifetimesB}
\end{figure}

In the bottom panel of Figure \ref{lifetimes} we show the effect of choosing different values of $f$ for pillar sizes $f r_{\textrm{I}}$ or using a constant pillar size. The results in section \ref{anal} reproduce the behaviour seen in the numerical models.  When the pillar size is $fr_{\textrm{D}}$ the globulette lifetime is a linear function of detach time and is nonlinear otherwise. In Figure \ref{lifetimesB} we compare the lifetimes given by two of our numerical models with those given by equation \ref{tl2}. Prior to the lifetime of the ionising source limiting that of the globulette confinement time, our analytic model gives results within 0.5\,Myr of the numerical result at all detach times.

\subsection{Uncertainties in the confinement timescale}
\label{uncertainty}
There are two significant approximations that we make in calculating the globulette confinement timescale that we have not yet discussed:  that globulettes initially travel at the velocity of the shell and that we have neglected rocket driven motion from photoevaporative outflows. 

\begin{figure}
	\includegraphics[width=8cm]{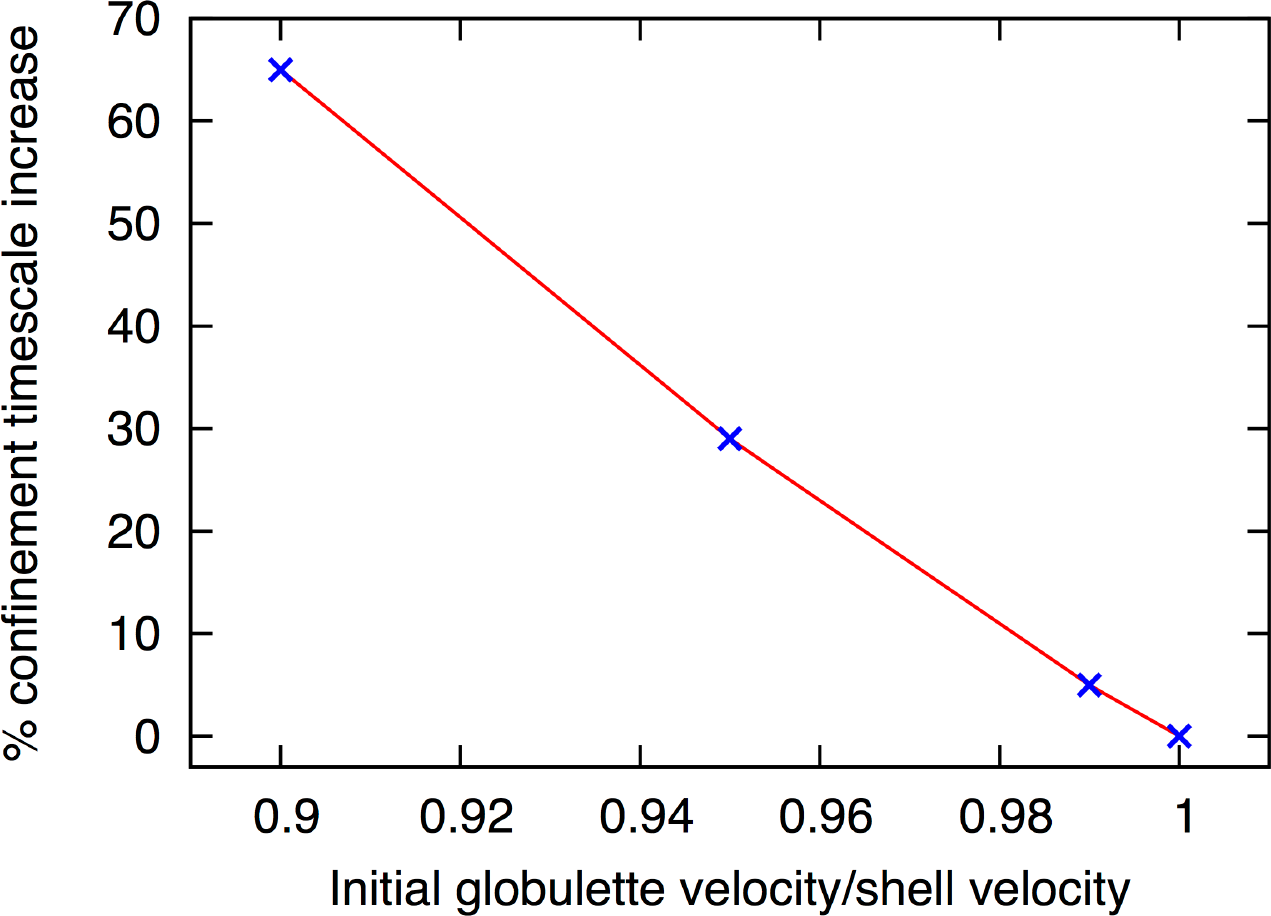}
		\caption{The effect upon the globulette lifetime of the initial globulette velocity being lower than the shell velocity. }
	\label{veleffect}
\end{figure}

If the globulettes are detached with lower velocities than the shell then they will remain confined to the H\,\textsc{ii} region for longer.  To quantify this we modify the initial globulette velocity for one of our numerical models and plot the percentage increase in the globulette confinement time as function of the fraction of the shell velocity that the globulette initially has in Figure \ref{veleffect}. Clearly lower initial velocities can increase the globulette confinement time significantly. We reiterate that \cite{2013A&A...555A..57G} found globulettes to be moving at the same velocity as the shell, but even a difference of only 5 per cent in the globulette velocity can increase the confinement time by 30 per cent.

Given that our model in which the globulettes are in pressure
equilibrium with the surrounding HII region provides an excellent fit
to the observational data (Figure 2), we do not expect that that the additional pressure
associated with photoevaporative flows from the exposed surface
of the globulette can be much larger than the ambient thermal pressure 
\citep[though][find a slight asymmetry in the density structure, with denser gas towards the OB association]{2013A&A...555A..57G}. In the case that these two pressures are similar in magnitude, the
one-sided application of photoevaporation pressure would result in
an outward acceleration of about 1\,km/s/Myr. From our numerical models, we find that typical globulettes have velocities of up to around 10\,km/s  and are confined up to 5\,Myr. Applying this acceleration to our numerical models we find that rocket driven motion can reduce the confinement timescale by up to a factor of $\approx2$ (though it is usually less than this at around 10-40 per cent). 

We therefore have two mechanisms that can significantly modify the confinement timescale acting to nullify one another. Determining which, if either, of these dominates requires radiation hydrodynamic simulations with photoionisation that self-consistently form pillars and globulettes. This kind of simulation would be at the forefront of modern numerical modelling of the expansion of H\,\textsc{ii} regions and is therefore beyond the scope of this paper. 

\section{Collisions with the shell}
In the event of a globulette colliding with the shell, we estimate the effect upon the globulette by comparing the kinetic energy  dissipated in the globulette with its  gravitational binding energy. The total kinetic
energy of the shell at the collision interface is
\begin{equation}
	K = \left[v-\dot{r_{\textrm{I}}}(t=t_{\textrm{L}})\right]^2\rho_{\textrm{s}}\pi R_{\textrm{G}}^2\Delta
	\label{K}
\end{equation}
where $\Delta$ is the shell width, $\rho_{\textrm{s}}$ the shell density and $\dot{r_{\textrm{I}}}$ the shell velocity. 

However  the quantity of energy dissipated within the globulette ($K'$) is
only a fraction $(\rho_{\textrm{s}}/\rho_{\textrm{G}})^{1/2}$ of this \citep{1991MNRAS.250..505S}.
Since the gravitational binding energy of the globulette is 
\begin{equation}
	U = \frac{3GM_{\textrm{G}}^2}{5R_{\textrm{G}}}.
\end{equation}
we then have that 
\begin{equation}
	\frac{K'}{U} = \frac{5\left[v-\dot{r_{\textrm{I}}}(t=t_{\textrm{L}})\right]^2\rho_{\textrm{s}}\pi R_{\textrm{G}}^3 \Delta}{3 G M_{\textrm{G}}^2} \sqrt{\frac{\rho_{\textrm{S}}}{\rho_{\textrm{G}}}}.
	\label{energyrat}
\end{equation}

We also need to consider the dissipation of energy through dust cooling. The cooling rate per unit volume from dust alone is
\begin{equation}
	\Lambda(T) = 4\kappa\rho\sigma T^4
\end{equation}
where $\kappa$ is the mean opacity (we use the Planck opacity).   Integrating this cooling rate per unit volume over the shell crossing timescale and the globulette volume gives the possible energy extracted during the shell crossing. Subtracting this energy lost gives

\begin{equation}
	\frac{K'}{U} = \sqrt{\frac{\rho_{\textrm{S}}}{\rho_{\textrm{G}}}}\frac{5R_{\textrm{G}}^3\Delta\pi}{3GM_{\textrm{G}}^2}\left[\rho_s(v-\dot{r_I})^2 - \frac{16}{3}\kappa\rho_{\textrm{G}} \sigma T_{\textrm{G}}^4 \frac{\Delta}{(v-\dot{r_I})}\right].
	\label{energyrat3}
\end{equation}
If $K'/U<1$ then the globulette is expected to survive the collision with the shell.

In order to evaluate the energy involved in the shell collision we also require a description for the evolution of the shell. If the shell is isothermal the density is given by the pre--shock density times the square of the Mach number in the pre--shock region
\begin{equation}
	\rho_{\textrm{s}} = \rho_{\textrm{a}}\left(\frac{\dot{r}_{\textrm{I}}}{c_{\textrm{a}}}\right)^2
	\label{rhoshellEVO}
\end{equation}
where $\rho_{\textrm{a}}$ and $c_{\textrm{a}}$ are the density and sound speed in the pre--shock gas respectively \citep{2007pafd.book.....C}.

{We calculate the mass in the shell as the difference between the total
mass of ambient gas intially contained within radius $r_{\textrm{I}} + \Delta$ and the
mass of ionised gas within $r_{\textrm{I}}$ (which is small in comparison). Since
the initial Stromgren radius, $r_{\textrm{s}}$,  defines a situation of ionisation
equilibrium for gas at density $\rho_a$, such equilibrum at radius
$r_{\textrm{I}}$ corresponds to an ionised gas mass of $4 \pi/3 \rho_{\textrm{a}} r_{\textrm{s}}^3 (r_{\textrm{s}}/r_{\textrm{I}})^{-3/2}$
(from equation (24)). Thus the shell mass and thickness are given by:}

\begin{equation}
	M_{\textrm{Tot}} = \frac{4\pi}{3}\left[(r_{\textrm{I}}+\Delta)^3 - r_{\textrm{I}}^3\right]\rho_{\textrm{s}} = \frac{4\pi}{3}\left[(r_{\textrm{I}}+\Delta)^3- r_{\textrm{s}}^3 (r_{\textrm{s}}/r_{\textrm{I}})^{-3/2}\right]\rho_{\textrm{a}}
\end{equation}
and 
\begin{equation}
	\Delta = \left[\left(\frac{r_{\textrm{I}}^3\rho_{\textrm{s}} - r_{\textrm{s}}^3 (r_{\textrm{s}}/r_{\textrm{I}})^{-3/2}\rho_{\textrm{a}}}{\rho_{\textrm{s}} - \rho_{\textrm{a}}}\right)^{1/3}-r_{\textrm{I}}\right].
	\label{deltaEVO}
\end{equation}
We evolve the shell parameters in our numerical calculations (section \ref{numcom}) and evaluate the collision once the globulette is coincident with the {shell}. 

Without dust cooling the collision and gravitational binding energies from equation \ref{energyrat3} are comparable. Those most susceptible to destruction are the smallest globulettes. To include dust cooling we calculate the Planck opacity using subroutines from the \textsc{torus} radiative transfer code  \citep{2000MNRAS.315..722H}.  We assume a dust--to--gas ratio of $10^{-2}$ and spherical, silicate dust grains that follow a standard interstellar medium power-law size distribution \citep{2003ARA&A..41..241D}. We assume a globulette temperature of 18\,K, following the observations of \cite{2007AJ....133.1795G}. {With these parameters for the lowest mass (and most easily destroyed) 0.1\,M$_J$ globulette} crossing the shell at 1\,km/s (the typical value from our subsequent numerical calculations) we find that 
\begin{equation}
	\frac{K}{U} \approx 2\left(\frac{\Delta}{\textrm{pc}}\right)\left[10^{10}\rho_s - 2\times10^{-5}\left(\frac{\Delta}{\textrm{pc}}\right)\right]\sqrt{\frac{\rho_{\textrm{S}}}{\rho_{\textrm{G}}}}.
	\label{energyrat4}
\end{equation}
Since $\left(\frac{\Delta}{pc}\right)$ will be not much less than order unity and the shell density is at most around $10^{5}$m$_{\textrm{H}}$/cm$^3$ the energy lost through dust cooling is always about 4--5 orders of magnitude higher than heating due to ram pressure. 
 Hence even the smallest globulettes will survive their encounter with the shell and escape into the wider ISM.  
 
 \section{Fate of globulettes in the wider ISM}
 
 Once in the wider ISM, dust cooling would continue to prevent the destruction of the globulette by the action of the surroundings. However, the external pressure could drop by an order of magnitude or more exterior to the H\,\textsc{ii} region so if globulettes are pressure confined within the H\,\textsc{ii} region then they will eventually dissipate. 
 
According to \cite{stan} it is impossible to find an exact analytic solution for a suddenly expanding sphere of gas in vacuum, however they calculate an {approximate} solution which is also given by  \cite{1967pswh.book.....Z} {where the globulette expansion velocity is
\begin{equation}
 \frac{dR_{\textrm{G}}}{dt} =  \frac{2}{\gamma-1}c_{\textrm{G}},
 \end{equation}
 and the globulette density evolves according to
 \begin{equation}
\rho = \frac{M_{\textrm{G}}}{R_{\textrm{G}}^3}\left(1 - \frac{R_{\textrm{G0}}^2}{R_{\textrm{G}}^2}\right)^{\alpha}
 \end{equation}
where
  \begin{equation}
{\alpha} = \frac{3-\gamma}{2(\gamma-1)}
 \end{equation}
and} $R_{\textrm{G0}}$, $c_{\textrm{G}}$, $\gamma$ are the initial globulette radius, sound speed in the globulette and the adiabatic index in the globulette respectively. 
 
 If we define the dissipation timescale as that at which the globulette density is equal to the ambient medium then
   \begin{equation}
t_{\textrm{Diss}} = \frac{1}{v}\left(\frac{M_{\textrm{G}}}{\rho_{\textrm{amb}}}\right)^{1/3}\left( 1-\frac{R_{\textrm{G0}}^2}{v^2 t^2} \right)^{\alpha/3}
 \end{equation}
 which has to be solved numerically unless in the limit of small $R_{\textrm{G0}}^2/(v^2t^2)$. Solving numerically and assuming $\gamma=1.4$ this gives dissipation timescales ranging from 30-300\,kyr over the globulette mass range observed in Carina. The dissipation time given in the aforementioned limiting case gives a value to within 0.5 per cent of the numerical result and so the simplified form is applicable.  Assuming a constant propagation velocity of 10\,km/s the largest globulettes might therefore be observable external to an H\,\textsc{ii} region out to a distance of up to 3\,pc (albeit increasingly dispersed). Of course this exact distance depends on the globulette properties and propagation velocity. Nevertheless globulettes might be observed using high spatial resolution molecular line observations at velocities similar to that of the shell. In and around Carina one would be able to resolve globulettes down to 1\,kAU at 230 GHz ($\approx 0.3''$) with an ALMA  baseline of 1\,km. 

We have found that the only way that globulettes might form brown dwarfs or free floating planets is if they are triggered to collapse as they traverse the shell. If they do form planets or brown dwarfs then they will contribute to the low mass tail of the local Initial Mass Function (IMF). In the central cluster NGC 2244 of the Rosette Nebula, recent estimates indicate a census of $\sim2000$ stars \citep{2005ApJ...625..242L,2008ApJ...675..464W}, whereas $145$ globulettes have been detected in the outskirts of the same cluster \citep{2007AJ....133.1795G}. The majority of these objects have masses below $30M_{\rm J}$. Considering the globulettes with $M<30M_{\rm J}$, if we assume they eventually collapse and form free floating planets and brown dwarfs they would constitute $\sim5$\,\% of the total number of stars in the cluster. This would imply a significant upturn in the IMF at masses below $30$\,M$_J$, compared with the IMF determined at higher (brown dwarf masses) by \cite{2012ApJ...748...14D} in the Orion nebular cluster. Globulettes could not however
possibly provide the very high numbers of free floating planets (outnumbering stars by a factor
two) claimed by \cite{2011Natur.473..349S}  using the MOA-II (Microlensing Observations in Astrophysics) survey \citep[see however][for a recent discussion on this topic]{Chabrier2014}.


\section{The distribution of globulettes in H\,\textsc{II} regions}

We find that globulettes within H\,\textsc{ii} regions will all eventually impact the shell. Since globulettes are not decelerated by ram pressure they soon travel faster than the shell. We therefore also find that globulettes detached from a pillar of size $L_{\textrm{p}}$ at the shell velocity remain within $L_{\textrm{p}}$ of the shell (this will not be the case if the initial globulette velocity is lower than the shell velocity). This is in qualitative  agreement with observations by \cite{2007AJ....133.1795G} and \cite{2013A&A...555A..57G} which find globulettes clustered near to the shell. We therefore suggest that any globulettes observed at larger distances interior to the shell either initially travel substantially slower than the shell or are formed via some other mechanism. For example, they could be the result of the irradiation of a turbulent medium, which has been shown to produce isolated (but much larger) globules in simulations by \cite{2012A&A...546A..33T} and \cite{2010ApJ...723..971G}. These two mechanisms can be distinguished by the motions of the globulettes. If they are detached from a shell then their propagation direction should correlate with the shell, whereas if they are the result of the irradiation of a turbulent medium simulations show that globulettes should have random velocities \citep{2012A&A...546A..33T}.




\section{Summary and conclusions}
We have developed a theoretical model describing many features of globulettes, including a description of their internal structure, stability, motions, their ability to survive collisions with the shell bounding an H\,\textsc{ii} region and their fate in the wider ISM.  We draw the following main conclusions from this work. \\

\noindent1. We demonstrate that globulettes are well described by pressure confined isothermal Bonnor-Ebert spheres (with very slowly varying density distribution until close to the globulette boundary). We have also shown that the observed $M\propto R^{2.2}$ relation for globulettes is explained by systematics, since there is a column density threshold below which components of the globulette will not be identified.   \\

\noindent2. We find that within the H\,\textsc{ii} region globulettes are very stable against collapse. Bonnor-Ebert spheres are stable against contraction if $\rho_{\textrm{c}}/\rho_{\textrm{o}} < 14.1$ and we find that this ratio is typically an order of magnitude lower for globulettes. \\

\noindent3. All globulettes will eventually impact the shell bounding the H\,\textsc{ii} region unless they are disrupted beforehand by, for example, a supernova or photoevaporation. This is because although  the globulette initially travels at the same speed as the shell {(we assume this based on observations)}, it experiences essentially zero deceleration due to ram pressure as the gas density in the H\,\textsc{ii} region is so low. \\

\noindent4. Given conclusions 2 and 3, the only way that globulettes might form brown dwarfs or free floating planets is if they are triggered to collapse as they traverse the shell at the boundary of the H\,\textsc{ii} region.\\

\noindent5. The time that a globulette remains confined to an H\,\textsc{ii} region (its {confinement timescale}) varies as the square root of the pillar size from which it detaches. If the pillar size scales as some constant fraction of the H\,\textsc{ii} region size then the globulette lifetime is a linear function of the detach time. If the pillar size is constant then the globulette {confinement timescale} is longer and a nonlinear function of detach time (approximately $t_{\textrm{L}} \propto t_{\textrm{D}}^{5/7}$). The {confinement timescale} is independent of the globulette mass (since ram pressure is negligible) and only varies with the ionising flux as $N_{\textrm{ly}}^{-1/6}$ {(though the rocket effect, which we do not directly include in this component of our analysis,  will increase the sensitivity to ionising flux)}. The lifetime is also sensitive to the initial globulette velocity, which we assume is initially equal to that of the shell, as observed by \cite{2013A&A...555A..57G}. {A globulette initially moving slower than the shell can be confined for much longer though. However we estimate that the rocket effect, which we do not directly include in our calculation, can decrease the confinement timescale by a similar factor. Radiation hydrodynamic calculations would be required to investigate these processes further. }\\

\noindent6. We find that because dust cooling can dissipate energy efficiently that all globulettes can survive their collision with the shell and escape into the wider ISM. Since the globulettes are pressure confined (see conclusion 2), they dissipate once they move into the lower pressure medium external to the H\,\textsc{ii} region, on timescale of order 30--300\,kyr, potentially allowing them to travel around 3\,pc. Small, high velocity clumps with similar motions to the shell might therefore be observable external to an H\,\textsc{ii} region in high spatial resolution molecular line observations (i.e. with ALMA).\\

\noindent7. Globulettes detached from a pillar of size $L_p$ at the velocity of the shell will remain within $L_p$ of the shell and will have velocities and trajectories similar to the shell. Globulettes at smaller radii could be formed through the irradiation of a turbulent medium and would have more random velocities \citep{2012A&A...546A..33T} or could have been detached at a velocity significantly lower than that of the shell, in which case their velocity should still be correlated with the shell. \\


\noindent 8. The number of globulettes observed in some H\,\textsc{ii} regions suggests that they could make a very significant
addition to the inventory of free floating planets/low mass brown dwarfs \textit{if} they were able to  collapse gravitationally.  Given their low Jeans numbers, this would however require some violent perturbative event.

\section*{Acknowledgments}
The authors wish to acknowledge the conference ``The Olympian Symposium on Star Formation (2014)'' where this work was triggered. We thank Pascal Tremblin, Tim Harries, Christopher Tout and, in particular, G\"osta Gahm for useful discussions. {We also thank the anonymous referee for their comments}. TJH is funded by the STFC consolidated grant {ST/K000985/1}. SF is funded by an STFC/Isaac Newton trust studentship.

\bibliographystyle{mn2e}
\bibliography{molecular}

\bsp

\label{lastpage}

\end{document}